\documentstyle[11pt,paspconf,epsf]{article}

\begin{document}
\def\hmin{\ifmmode{h^{-1}}\else{$h^{-1}$}\fi}
\def\kms {\ifmmode{{\rm km\,s}^{-1}}\else{km\,s$^{-1}$}\fi}
\def\msun{\ifmmode{{\rm M}_\odot}\else{${\rm M}_\odot$}\fi}
\def\Mpc{{\rm Mpc}}
\def\hminMpc3{\ifmmode{h^3{\rm Mpc}^{-3}}\else{$h^3{\rm Mpc}^{-3}$}\fi}

\title{Projection Effects in the Abell Catalogue}
\author{M.P. van Haarlem}
\affil{Department of Physics, University of Durham, South Road, Durham
DH1 3LE, UK}

\begin{abstract}
We investigate the influence that projection effects have on the
completeness of the the Abell cluster catalogue in the context of an
$\Omega_0=1$ CDM Universe.  We construct a galaxy catalogue and
identify clusters using an algorithm that mimics Abell's method.  Over
30\% of Richness Class $\geq$1 clusters have been selected despite not
meeting the required intrinsic (3D) richness criterion. Velocity
dispersions are shown to have been systematically overestimated.
\end{abstract}


\section{Introduction}

The abundance and masses of clusters of galaxies are a very sensitive
probe of both the spectrum of initial fluctuations and of the
cosmological parameter $\Omega_0$.  The clusters in most catalogues
available at present have been detected as an overdensity of galaxies
on photographic plates.  However, the 2D nature of this selection
procedure leads to contamination by both foreground and background
galaxies.  The incompleteness this causes is likely to have serious
consequences for measures such as the cluster-cluster correlation
function and the distribution of cluster velocity dispersions.

The most widely used optically selected (2D) cluster catalogues are
those compiled by Abell (1958) and ACO (Abell et al. 1989).  Recent
machine based surveys (Dalton et al. 1992; Lumsden et al. 1992) have
yielded catalogues of only small regions of the sky (10-20\%, compared
with the all-sky ACO catalogue) but should suffer less from projection
effects.  At present though, most studies that use clusters are still
based on the Abell and ACO catalogues; for example, the recently
completed ESO Key Programme on Rich Clusters of Galaxies (Katgert et
al. 1996). This study has shown that many Abell clusters have more than
a single system along the line of sight.  By taking over 50 redshifts
for a large number of clusters, they have established that such
projection effects are fairly common.  Judging from the data presented
by Katgert et al. (1996), it appears as though a significant fraction
(perhaps as high as 30\%) of Richness class $\geq$1 clusters may be
seriously affected by projection effects.

It is obvious that a careful investigation of the completeness of
optical cluster catalogues is warranted.  We report here on the first
results of such a study, using artificial galaxy catalogues.  Similar
techniques were used by Frenk et al. (1990) and White (1991,1992).  A
more comprehensive account can be found in two forthcoming papers (van
Haarlem, Frenk \& White 1996; van Haarlem \& Frenk 1996).

\section{Construction of a Galaxy and Cluster Catalogue}

The results discussed in this paper are based on a set of 8 Standard
CDM N-body simulations described more fully by Eke et al. (1995) and
performed with an adaptive P$^3$M code using $128^3$ particles in an
$L_{box}=256 \hmin\Mpc$ comoving periodic box.  At the output time used
for the galaxy catalogue $\sigma_8$, the $rms$ amplitude of mass
fluctuations in a sphere with radius 8 \hmin\Mpc, had a value of 0.63.
Only a single epoch was used, in order to concentrate on the
differences caused by the selection of clusters in 2D and not by the
inherent uncertainties in the galaxy formation process.

\subsection{Constructing Galaxy Catalogues}

There is no satisfactory method that allows us to construct a large
catalogue of galaxies while treating the dissipative processes of star
and galaxy formation and their evolution in a representative way.
However, such a catalogue is clearly required, if we are to select
clusters in a manner similar to the way they are found on photographic
plates.  We have chosen to use the peak-background split technique to
make a selection of the total number of particles available, and
identify them with individual galaxies.  The idea behind this process
is that galaxies will form at the sites where the initial density field
exceeds a threshold on a scale that we associate with that of an
individual galaxy.  Because our simulations cover a large volume, the
initial resolution is only 4{\hmin\Mpc}, which is somewhat greater than
the scale $r_s$ of a galaxy (on average the mass contained within a
top-hat with a radius $r_s=0.54\hmin\Mpc$ is of the order of
$10^{12}\msun$).  The peak-background split technique uses the
machinery developed by Bardeen et al. (1986) to predict the number
density of peaks on a scale $r_s$ in a Gaussian Random Field which we
can only resolve on a scale $r_b>r_s$.  Further details can be found in
White et al. (1987) and in van Haarlem et al. (1996).  We first smooth
the initial density field, using a sharp $k$-space filter to remove all
power below $r_b$ which eliminates artificial correlations between the
smooth field and the peaks we are looking to find.  Using a value of
$r_b$=8.75{\hmin\Mpc}, we calculate the conditional probability that
the initial density field exceeds a local threshold for galaxy
formation, given its actual value found in the smoothed field. The
procedure is applied to all the points in the density field, sampled at
the initial locations of the particles on the grid.  A final (volume
limited) catalogue is compiled by taking a random sample from the
particles, while considering the probability computed above.  The total
number of galaxies is determined independently by requiring that the
luminosity density of our model catalogue is consistent with recent
determinations of luminosity function parameters (e.g. Loveday et al.\,
1992; Marzke et al.\, 1994).  The value of $\rho_L = 0.0176
L_*\hminMpc3$ also reproduces the observed abundance of Abell clusters
($\sim 8\times 10^{-6} \hminMpc3$; Bahcall \& Soneira 1983) when we use
the method described below to identify the clusters.  Where necessary,
we have assumed that the galaxies follow a Schechter luminosity
function
\begin{equation}
\phi(L)dL = N_* L^{-\alpha} \exp(-L)/\Gamma(2-\alpha)dL,
\end{equation}
where $L$ is expressed in units of $L_*$, and $\Gamma$ is the Gamma
function.  Apart from matching the observed number of Abell clusters,
the two point correlation function of the galaxies in our 
catalogue also agrees well with the observed $\xi_{gg}$.

\subsection{Cluster Selection}

Each volume limited catalogue was projected onto a plane along the
three coordinate axes, thus producing three different 2D galaxy
catalogues from each simulation.  A friends-of-friends group finding
algorithm with a linking length 30\% of the mean intergalaxy separation
was then applied to the 2D catalogue.  The list of
groups it produces forms the starting point of the search for clusters
using a procedure that mimics Abell's selection criteria closely.

Abell defined a rich cluster as an enhancement of galaxies on the
Palomar Sky Survey plates.  To qualify as a cluster, the number of
galaxies within the Abell radius ($r_a=1.5\hmin\Mpc$) from the proposed
cluster centre had to exceed $n_a$, after subtracting the background
count.  These galaxies were constrained to the magnitude interval
between $m_3$ and $m_3 + 2$, where $m_3$ is the apparent magnitude of
the third brightest galaxy.  For the lowest richness class that is
often assumed to be reasonably complete, $n_a$ has to exceed the
background count by at least 50 galaxies (Abell richness class $R$=1).
The relation between $n_a$ and the luminosity density of each catalogue
follows from the assumed functional form of the luminosity function and
this richness criterion.  As is shown in more detail in van Haarlem et
al. (1996) this assumption leads to the following relation
\begin{equation}
n_a = \frac{N_*}{\Gamma(2-\alpha)} \int\limits_{0.1585
      \overline{L}_3}^{\overline{L}_3} L^{-\alpha} \exp(-L) dL.
\end{equation}
Here, $N_* L_*$ represents the total luminosity projected within $r_a$
and $\overline{L}_3$ is the median luminosity of the third brightest
galaxy.  The 50 galaxies that are needed for an $R=1$ cluster
correspond to $60L_*$, which is achieved when the total catalogue
contains $N_{gal}=\rho_L n_a V / N_*$ galaxies.  For the present set of
simulations $N_{gal}\sim 3.4\times 10^5$.  Since the number of galaxies
that are found in clusters is only a small fraction of the total number in
each volume we may assume that on average the fore- and background 
contribution is
directly proportional to the volume projected onto the cluster, i.e. a
cylinder with a radius $r_a$ and length $L_{box}$.  Within an Abell
radius we therefore expect, on average, a contamination of 27 galaxies,
in addition to the 50 that belong to the cluster.  Each of the groups
identified using the friends-of-friends method in the 2D galaxy
catalogue was checked.  The poorer of a pair of overlapping clusters
(projected separation $< 2 r_a$) was removed.  On average 139 clusters
were found in each catalogue ($\sim 8\times 10^{-6} \hminMpc3$).

One final preliminary is the translation of the 60$L_*$ that represents
an $R=1$ cluster in a 2D catalogue, to an intrinsic (3D) luminosity.  This
value can be obtained by deprojection of a 2D profile, and comparing
the integrated mass within a cylinder (2D) and sphere (3D).  For a
power-law density profile $\rho(r)\propto r^{-\gamma}$, with
$\gamma\sim 2.2$ as found by Lilje \& Efstathiou (1988), the ratio turns
out to be 0.72.  We will therefore consider clumps of galaxies that
have a luminosity greater than 43$L_*$ as 3D Abell clusters.

\begin{figure}
\centering
\centerline{\epsfysize=7.0truecm
\epsfbox[60 200 550 650]{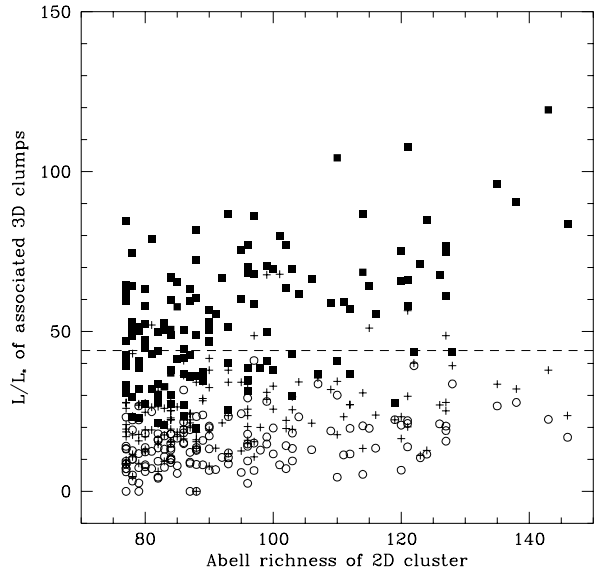}
\message{Figure 1}}
\caption{The luminosity of the three most luminous system along the
line of sight to each cluster in one of the cluster catalogues as a 
function of the apparent 2D galaxy count. See text for details}
\end{figure}

\section{Completeness of the Cluster Catalogue}

In order to assess the success of the cluster selection technique in
identifying real clusters while rejecting projection effects, we have
computed the luminosity associated with all clumps of particles in the
original N-body output.  Once again we used a friends-of-friends
algorithm, but now with a small linking length (10\% of the mean
interparticle separation) in order to pick up small groups.  In
Figure~1 we show the results for one of our projected catalogues. Along
the horizontal axis we have plotted the number of galaxies counted
within a projected Abell radius (the background count has not been
subtracted, hence an $R=1$ cluster corresponds to $n\geq 77$).  The
total luminosity within a sphere with radius $r_a$ centred on the three
most luminous groups found along the line of sight to each cluster is
plotted along the vertical axis.  The main cluster is shown as a filled
square, the second and third most luminous group as a cross and circle
respectively.  The dashed line shows the 43$L_*$ that we derived above.
Averaged over all 24 catalogues, a significant fraction (34$\pm$6\%) of
all clusters do not have a single clump along the line of sight which
has a luminosity in excess of 43$L_*$.  We also found that 32$\pm$5\%
of the total number of groups that are brighter than 43$L_*$ were not
associated with $R\geq$1 clusters.  This last effect is due to a low
background count, which has effectively led to an overestimate of the
fore- and background contribution.  The overall conclusion we come to
is that approximately a third of all $R\geq$1 clusters must be replaced
by clusters that have not been picked out by the selection procedure.
Although lowering the threshold, or calculating a local background
value would presumably lead to a higher detection rate, it would also
increase the number of phantom clusters.

\section{Velocity Dispersions}

Through the virial theorem, the velocity dispersions of clusters
provide a means of deriving the mass distribution of clusters, which is
known to be a powerful discriminant between different cosmological
scenarios (White, Efstathiou \& Frenk 1993; Bahcall \& Cen 1993).  A
useful way of looking at the velocity dispersions is by plotting their
cumulative distribution function $n(>\sigma_{v,los})$, the number
density of line of sight velocity dispersions.  Projection effects will
have a tendency to increase the high $\sigma_v$ tail of the
distribution, because small clumps in the vicinity of the cluster are
difficult (if not impossible) to remove. In Figure 2 we show the effect
of 3 different methods of calculating $\sigma_v$.  Firstly, we compute
$\sigma_v$ in an analogous manner to the way most observational samples
have been treated.  Using the redshifts of all the galaxies within a
cylinder with radius $r_a$ centred on the cluster, we identify the peak
of the histogram. All galaxies more than 4000{\kms} from this peak are
removed, and an iterative 3$\sigma$-clipping technique is applied to
the remaining data.  The resulting $n(>\sigma_{v,los})$ is shown as the
filled squares in Figure 2.  It is quite noticeable that there appears
to be a significant tail in the distribution, extending to at least
2000{\kms}.  By using a smaller aperture, the cross-section for
possible projection effects is sharply reduced. One also notices the
effect on $\sigma_v$.  We have applied the same algorithm described
above to all the galaxies within $r_a/3=0.5\hmin\Mpc$, and find a
marked reduction in the number of high velocity dispersion clusters
(open squares).  Finally, the triangles in Figure 2 use the iterative
scheme developed by den Hartog \& Katgert (1995).  Within an aperture
of 1.0{\hmin\Mpc}, they apply a radially varying criterion for the
elimination of interlopers.  At each projected radius, they determine
the extreme values of the radial component of both the infall and
circular velocity, and reject all galaxies that exceed these extremes.
In each case they use conservative limits in order not to reject too
many galaxies.  This leads to a further decrease in the amplitude of
the high-$\sigma_v$ tail (triangles in Figure 2).  However, none of
these techniques is capable of approaching the intrinsic (3D) velocity
dispersions, obtained directly from the dark matter particles within a
sphere with radius $r_a$.  The distribution of $\sigma_v$ of all dark
matter clumps is shown as the solid line.  This curve also reveals that
our richness selected catalogue is very incomplete below
$\sigma_v\sim1000{\kms}$.  If these factors are taken into account,
then a comparison of velocity dispersions determined from clusters
taken directly from N-body simulations and compared with observations
must lead to serious errors.  Bahcall \& Cen's (1993) rejection of the
$\Omega_0=1$ CDM model on the basis of the cluster mass function
therefore seems somewhat premature.
 
\begin{figure}
\centering 
\centerline{
\epsfxsize=6.5truecm 
\epsfbox[75 225 500 600]{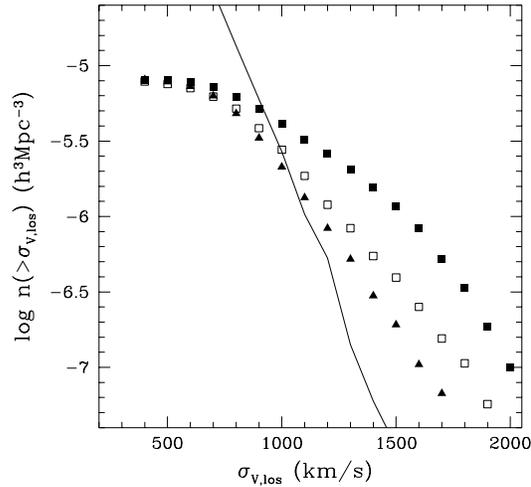}
\message{Figure 2}}
\caption{The cumulative distribution of cluster velocity dispersions
based on all synthetic cluster catalogues. See text for details.}
\end{figure}

\section{Conclusions}
We have shown that, in the context of an $\Omega_0=1$ CDM Universe,
contamination is a serious problem that leads to around one-third of
clusters being misclassified as an $R\geq$1 Abell cluster, while a
similar fraction is not recognised.  Projection effects also influence
the velocity dispersions that one can estimate from observations.
Even the most elaborate techniques of removing interlopers, are
not able to recover the true 3D velocity dispersion distribution of
dark matter clumps.  

\acknowledgments
\begin{sloppypar}
I thank my collaborators, C.S. Frenk and S.D.M. 
White and acknowledge financial support from an EU HCM fellowship.
\end{sloppypar}


\begin{references}
\reference Abell G. O., 1958, ApJS, 3, 211
\reference Abell G. O., Corwin H. G., Olowin R. P., 1989, ApJS, 70, 1
\reference Bahcall N. A., Cen R., 1993, ApJ, 407, L49
\reference Bahcall N. A., Soneira R. M., 1983, ApJ, 270, 20
\reference Bardeen J. M., Bond J. R., Kaiser N., Szalay A. S., 1986,
ApJ, 304, 15
\reference Dalton G. B., Efstathiou G., Maddox S., Sutherland W. 1992, 
ApJ, 390, L1
\reference den Hartog, R. H., Katgert P., 1995, MNRAS in press.
\reference Eke V. R., Cole S., Frenk C. S., Navarro J. F. N., 1995, 
MNRAS, In Press
\reference Frenk C. S., White S. D. M., Efstathiou G., Davis M., 1990, 
ApJ, 351, 10
\reference Katgert P., Mazure A., Perea J., den Hartog R. et al., 1995,
A \& A  In Press
\reference Lilje P. B., Efstathiou G., 1988, MNRAS, 231, 635
\reference Loveday J., Peterson B., Efstatiou, G. and Maddox, S., 1992, 
ApJ, 390, 338 
\reference Lumsden S. L., Nichol R. C., Collins C. A., Guzzo L., 1992, 
MNRAS, 258, 1
\reference Marzke R. E., Huchra, J. P., Geller M. J., 1994, ApJ, 428,43
\reference White S. D. M., Briel U. G., Henry J. P., 1993, MNRAS, 261, L8 
\reference White S. D. M., Efstathiou G., Frenk C. S., 1993, 
MNRAS, 262, 1023
\reference White S. D. M., Frenk C. S., Davis. M., Efstathiou G., 1987,
ApJ, 313, 505
\reference White S. D. M., 1991, Large-Scale Structures and Peculiar
Motions in the Universe (Eds. Latham D. W. \& Da Costa L.N.) ASP, 285
\reference White S. D. M., 1992, Clusters and Superclusters of Galaxies
(Ed. Fabian A. C.) Kluwer, 17
\end{references}
\end{document}